\documentclass{article}
\newcommand{\R}{\mathbb{R}}
\newcommand{\C}{\mathbb{C}}
\newcommand\Cl{\mathfrak{Cl}}
\newcommand\Id{\mathbf 1}
\newcommand{\Dir}{\mathcal{D}}
\newcommand{\pd}[1]{\frac{\partial}{\partial{#1}}}
\newcommand\eq[1]{Eq.~(\ref{#1})}
\newcommand\nPsi{\mathsf{\Psi}}
\newcommand{\LA}{\mathit\Sigma}
\newcommand{\Ext}{\Lambda}
\newcommand{\ClLGr}{\mathfrak{C}_\mathcal{L}}
\newcommand{\ClRGr}{\mathfrak{C}_\mathcal{R}}
\newcommand{\ExCl}{\mathtt{\Lambda}_\mathfrak{C}}
\newcommand{\rgact}[1]{\overleftarrow{#1}}

\usepackage{amsfonts}
\usepackage{exscale}

\title{Dirac Spinors and Representations of GL(4) Group in GR}
\date{}
\author{Alexander Yu.\ Vlasov}
\begin{document}
\maketitle
\begin{abstract}
  Transformation properties of Dirac equation correspond to Spin(3,1)
representation of Lorentz group SO(3,1), but group GL(4,$\R$) of general
relativity does not accept a similar construction with Dirac spinors.
On the other hand, it is possible to look for representation of
GL(4,$\R$) in some bigger space, there Dirac spinors are formally situated
as some ``subsystem.'' In the paper is described construction of such
representation, using Clifford and Grassmann algebras of 4D space.
\end{abstract}

\section{Introduction}

Let as consider the Dirac equation \cite{DauIV,ClDir}
($x_0=ct$, $\hbar=1$, $c=1$)
\begin{equation}
\Dir\psi = m \psi, \quad \Dir \equiv \sum_{\mu=0}^3 \gamma_\mu  \pd{x_\mu},
\label{Dir}
\end{equation}
where $\psi \in \C^4$ and $\gamma_\mu \in Mat(4,\C)$, $\mu = 0,\cdots,3$ are
four Dirac matrices with usual property
\begin{equation}
 \gamma_\mu \gamma_\nu + \gamma_\nu \gamma_\mu = 2 g_{\mu\nu} \Id,
\label{acom}
\end{equation}
where $g_{\mu\nu}$ is Minkowski metric.

More generally, \eq{acom} corresponds to definition of Clifford algebra
$\Cl(g)$ for arbitrary quadratic form $g_{\mu\nu}$ and dimension. Dirac
matrices are four generators for
(complexified) representation of $\Cl(3,1)$ in space of $4 \times 4$ matrices.
For general universal Clifford algebra with $n$ generators $\dim\Cl(n)=2^n$.

It is possible to write Dirac equation for arbitrary metric, but covariant
transformation between two solutions $\psi$  exists only for isometries
of coordinates $A$: $g(Ax,Ay)=g(x,y)$ for given fixed metric $g$. Such
isometries produce some subgroup of GL(4,$\R$), isomorphic to Lorentz group
(for diagonal form, i.e. Minkowski metric, it is usual Lorentz group).
It is briefly recollected in Sec.~\ref{Sec:Spin}.
A matrix extension of Dirac equation is considered in Sec.~\ref{Sec:Ext}.
It is analyzed using Clifford (Sec.~\ref{Sec:Cliff}) and Grassmann
representation (Sec.~\ref{Sec:Grass}) of such equation. Main proposition
about representation of GL(4,$\R$) is formulated at end of
Sec.~\ref{Sec:Grass} and discussed with more details in Sec.~\ref{Sec:Expl}.

\section{Spinor Representation of Lorentz Group}
\label{Sec:Spin}

Any transformation $v'= Av$ of  coordinate system corresponds to new set
of gamma matrices:
\begin{equation}
 \gamma_\mu' = \sum_{\nu=0}^3 A_\mu^\nu \gamma_\nu,
\label{Agam}
\end{equation}
but only for isometries $A$ \eq{Agam} may
be rewritten as internal isomorphism of algebra:
\begin{equation}
 \gamma_\mu' = \LA_{(A,g)} \gamma_\mu \LA_{(A,g)}^{-1},
\quad \mbox{(there is no sum on $\mu$)}
\label{LgL}
\end{equation}
where matrix $\LA_{(A,g)}$ is same for any $\gamma_\mu$ and depends
only on transformation $A$ and metric $g$. For Minkowski metric $g$,
it is usual form of spinor $2\to 1$ representation of Lorentz group.

Similar with general case, spin group is implemented here as subset of
Clifford algebra. More precisely, it is subset of $\Cl_e$, {\em i.e.}
even subalgebra of Clifford algebra generated by all possible products
with even number of generators \cite{ClDir}. In particular case of Lorentz
group SO(3,1), it is Spin(3,1) group and isomorphic with SL(2,$\C$),
group of $2\times 2$ complex matrices with determinant unit.

To demonstrate transformation properties of spinor $\psi$, it is enough
to rewrite \eq{Dir} as
\begin{equation}
(\LA\Dir\LA^{-1})\LA\psi = m \LA\psi.
\label{LDirL}
\end{equation}
The \eq{LDirL} also shows, why \eq{Agam} with general $A \in \rm GL(4,\R)$
may not correspond to such kind of covariant transformation
$\psi(x) \mapsto \LA_A \psi(A x)$.

\section{An Extension of Spinor Space}
\label{Sec:Ext}

One way to overcome problem with representation of general coordinate
transformation of GL(4,$\R$) group may be extension of linear space $\C^4$
of Dirac spinors.\footnote{Say, two-component spinors
may not represent $P$-inversion, but extension from $\C^2$ to $\C^4$,
from Pauli to Dirac spinors resolves this problem \cite{DauIV}.}

There is also more technical way to explain idea of extension. It is not
possible to bulid some specific symmetry between two solutions of Dirac
equation with different metric, but may be it is possible to construct
a new one using {\em a few} solutions?

Let us consider instead of $\psi \in \C^4$ some $4 \times 4$ complex matrix
$\nPsi$ and write equation
\begin{equation}
\Dir\nPsi = m \nPsi, \quad \Dir \equiv \sum_{\mu=0}^3 \gamma_\mu  \pd{x_\mu},
\quad \nPsi \in Mat(4,\C).
\label{nDir}
\end{equation}
Formally it may be considered as
set of four usual Dirac equations for each row of matrix $\nPsi$.

Maybe such extension from 4D to 16D space is not minimal, but it is
appropriate for purpose of present paper, {\em i.e.} for representation
of transformation properties of such equation with respect to
GL(4,$\R$) group of coordinate transformations. For 16D space all possible
linear transformations may be represented by 256D space, but application
of algebraic language may simplify consideration. The construction uses
both Clifford and Grassmann algebras of 4D space and described below.

\section{Clifford Algebra}
\label{Sec:Cliff}

Complexified Clifford algebra of Minkowski quadratic form is isomorphic
with space of $4 \times 4$ complex matrices. So it is reasonable to try
consider \eq{nDir} as equation on Clifford algebra
$\gamma_\mu,\nPsi \in \Cl(3,1)$. Really, as it will be shown below, complete
and rigour solution of discussed problem with GL(4,$\R$) representation
may not be based {\em only} on Dirac equation on Clifford algebra, but
this construction discussed here, because provides an essential step.

If to consider \eq{nDir} as equation on Clifford algebra, then initial
Dirac equation may be compared first with restriction of such equation on
{\em left ideal} of Clifford algebra.

Left ideal of algebra $\mathcal A$ by definition \cite{SLang}
is linear subspace $\mathcal I \subset \mathcal A$ with property
$\mathcal{A I} \subset \mathcal I$, {\em i.e.} any element of algebra after
multiplication on element of an ideal produces again element of the ideal.
Simplest example of left ideal in matrix algebra is set of matrices
$\mathsf M_\psi$ with only one nonzero column $\psi$ and it provides reason
for consideration of Dirac equation on such ideal as an analogue of usual
case \eq{Dir}.

It was already discussed above, that spin group may be naturally implemented
as subspace of Clifford algebra, {\em e.g.} transformations of $\nPsi$
are also elements of Clifford algebra, $\LA \in \Cl(3,1)$.
On the other hand, construction with ideals of Clifford algebra have some
problems with interpretations of symmetries of Dirac equation, necessary for
purposes of given paper. Even for Lorentz transformation, simply represented
as isomorphisms of Clifford algebra \eq{LgL}, instead of \eq{LDirL} we must
have {\em the same} transformation law for all elements of algebra
\begin{equation}
(\LA\Dir\LA^{-1})\LA\nPsi\LA^{-1} =
 m \LA\nPsi\LA^{-1}.
\label{LPsiL}
\end{equation}

Really it makes consideration a bit more difficult, but does not change it
much. Resolution of the problem, is additional symmetry of \eq{nDir}:
if some $\nPsi$ is solution of \eq{nDir}, then $\nPsi R$ is also solution,
for arbitrary element $R$ of Clifford algebra.

In such a case right multiplication on $\LA^{-1}$ does not changes
anything and may be ignored. The same property of equation requires
consideration not only element $\mathsf M_\psi$ of some left ideal, but also
all $\mathsf M_\psi R$, {\em i.e.} matrices with columns proportional to same
vector.

If $\psi \in \C^4$ is initial
vector (spinor, solution of Dirac equation), and $\alpha \in \C^4$ is
arbitrary vector of coefficients, then any matrices with proportional
columns may be expressed as $M_{ij}=\psi_i\alpha_j$.

So instead of left ideals discussed above it is necessary to consider
matrices
\begin{equation}
 M_{ij}=\psi_i\alpha_j,
 \quad M = \psi \alpha^{T} \equiv \psi \otimes \alpha.
\label{bipsi}
\end{equation}
For arbitrary matrices $L,R$
\begin{equation}
  LMR = L\,(\psi \otimes \alpha)\,R = (L\psi) \otimes (R^{T} \alpha),
\label{LMR}
\end{equation}
so multiplication saves ``the product structure,'' but it is not an ideal,
because space of such matrixes is not {\em linear subspace}, {\em e.g.}
sum of elements does not necessary may be presented as tensor product
of two vectors like \eq{bipsi}. Really linear span of the ``singular'' space
coincides with whole algebra.

On the other hand, fixed $\psi$ corresponds to a linear subspace,
{\em right ideal} $\mathcal R_{\psi}$ of the algebra. It is similar with
interpretation of physical solution of usual Dirac equation \eq{Dir} as
a ray in Hilbert space.

But such construction still not
produce covariant transformation of Dirac equation in
matrix form \eq{nDir} with respect to general element of GL(4,$\R$).
It is necessary to use slighly different construction
with Grassmann algebra described in next section.

\section{Grassmann Algebra}
\label{Sec:Grass}

Formally Grassmann (or exterior) algebra $\Ext_n$ is defined by $n$
generators $d_i$, associative operation denoted as $\wedge$ and property
\begin{equation}
 d_\mu \wedge d_\nu + d_\nu \wedge d_\mu = 0.
\label{gras}
\end{equation}
So $\dim\Ext_n=2^n$, similarly with Clifford algebra.
Linear subspace of Grassmann algebra generated by $\wedge$-product of $k$
different elements $d_i$ are usually denoted as $\Ext_n^k$ (``$k$-forms'').
For convenience here is used complex Grassmann algebra.

On the other hand, Clifford algebra, Dirac operator and spin group also may
be expressed using Grassmann algebra \cite{ClDir}.
Let us consider Grassmann algebra $\Ext_n$ of $n$-dimensional vector
space $V$ and metric $g$ on $V$. Algebra of linear transformations of
Grassmann algebra is denoted here as $\mathcal L(\Ext_n)$ and for
any vector $v\in V$, it is possible to construct linear transformations
\cite{ClDir} $\delta_v,\delta^\star_v \in \mathcal L(\Ext_n)$:
\begin{equation}
 \delta_v : v_1\wedge\cdots\wedge v_k
 \mapsto v \wedge v_1\wedge\cdots\wedge v_k,
\label{bnd}
\end{equation}
\begin{equation}
 \delta^\star_v : v_1\wedge\cdots\wedge v_k \mapsto
 \sum_{l=1}^k (-1)^l g(v,v_l)\, v_1\wedge\cdots \not\!v_l\cdots\wedge v_k,
\label{cobnd}
\end{equation}
where ${\not\!v_l}$ means, that term $v_l$ must be omitted.
Let $v_i$ is basis of $V$, then operators
\begin{equation}
  \hat\gamma_i = \delta_i + \delta^\star_i\qquad
 (\delta_i \equiv\delta_{v_i},\ \delta_i^\star \equiv\delta_{v_i}^\star)
\label{gamLGr}
\end{equation}
satisfy usual relations with anticommutators \eq{acom} for $n$ generators of
Clifford algebra with quadratic form $g$, and so $\Cl(g)$ may be represented
as subspace of $\mathcal L(\Ext_n)$. Let us denote this representation
\begin{equation}
 \ClLGr : \Cl(n) \to \mathcal L(\Ext_n).
\label{ClGr}
\end{equation}

\medskip

Let us consider also canonical isomorphism of Grassmann and Clifford algebras
$\ExCl : \Ext_n \to \Cl(n)$ {\em as linear spaces} defined on basis as
\begin{equation}
 \ExCl : d_{i_1} \wedge \cdots \wedge d_{i_k}
 \mapsto \gamma_{i_1} \cdots \gamma_{i_k},
\quad i_1< i_2 < \cdots < i_k,
\label{ExtToCl}
\end{equation}
together with inverse one $\ExCl^{-1}$.

It is possible to express basic property of formal
constructions above as
\begin{equation}
 \ExCl^{-1}(\mathsf{LM}) = \ClLGr(\mathsf L)\bigl(\ExCl^{-1}(\mathsf M)\bigr),
 \quad \mathsf{L,M} \in \Cl.
\end{equation}

It should be mentioned, that using Hodge operator
$\star : \Ext_n^k \to \Ext_n^{n-k}$ it is possible to
write $\delta^\star_v = \star\,\delta_v \star$. By using
dual Grassmann operation $\vee = \star \wedge \star$, it is possible to
simplify \eq{bnd}, \eq{cobnd}
\begin{equation}
 \delta_v (\omega) = v \wedge \omega,
 \quad \delta^\star_v (\omega) = v \vee \omega, \quad \omega \in \Ext_n.
\end{equation}

Similarly, it is possible to introduce right actions
\begin{equation}
 \rgact\delta_v (\omega) = \omega \wedge v,
 \quad \rgact\delta^\star_v (\omega) = \omega \vee v, \quad
  \rgact{\hat\gamma}_i = \rgact\delta_i + \rgact\delta^\star_i,
\end{equation}
and right representation $\ClRGr$ of Clifford algebra
in $\mathcal L(\Ext_n)$ with property
\begin{equation}
 \ExCl^{-1}(\mathsf{LMR}) =
 \Bigl(\ClLGr(\mathsf L)\circ\ClRGr(\mathsf R) \Bigr)
 \bigl(\ExCl^{-1}(\mathsf M)\bigr),
 \quad \mathsf{L,M,R} \in \Cl.
\end{equation}

\medskip

The Dirac equation also has natural representation here \cite{ClDir}.
It is possible to express Dirac operator using exterior differential for
forms $d$ and its Hodge dual $d^\star$
\begin{equation}
 d = \sum_{i=1}^n \delta_i \pd{x_i},\quad
 d^\star = \sum_{i=1}^n \delta_i^\star \pd{x_i},\quad
 \Dir = d + d^\star.
\label{HodgeDir}
\end{equation}

\medskip

It is also possible to use standard representation of spin group via
left representation $\ClLGr$ of Clifford algebra \eq{ClGr}. Usually it
is restricted on spaces of odd and even forms \cite{ClDir}.

\smallskip

On the other hand, unlike of Clifford space, Grassmann space also accept
external GL(4,$\R$) isomorphism induced by extension of transformation
\begin{equation}
 d_\mu' = \sum_{\nu=0}^3 A_\mu^\nu d_\nu.
\label{Adiff}
\end{equation}
to general form $d_{i_1} \wedge \cdots \wedge d_{i_k}$.
Existence of such kind of isomorphism is common property of antisymmetric
forms, and Grassmann algebra may be represented as exterior algebra of the
forms.

Representation of Dirac operator in form \eq{HodgeDir} ensures
proper transformation property. It is clear for exterior differential $d$,
and for $d^\star$ it is also so, because terms with $g(v,v_l)$ in
\eq{cobnd} are transformed to $g(Av,Av_l)$, {\em i.e.} in agreement with
change of metric $g'=A^T g A$ and also demonstrate desired
property of $\Dir = d + d^\star$ with respect to map $\ClLGr$
and \eq{Agam}.

So result of present paper may be formulated as:

{\bf Proposition:}
{\em Transformation properties of solution of Dirac equation in matrix form
\eq{nDir} with respect to group {\rm GL(4,$\R$)} due to isomorphism
$$\ExCl : \Ext_4 \longrightarrow \Cl(3,1) \cong Mat(4)$$
corresponds to standard transformation properties of exterior algebra of
antisymmetric forms on tangent space with respect to general linear
coordinate transformations from {\rm GL(4,$\R$)}.}

\section{Further Discussion}
\label{Sec:Expl}

The main proposition needs for some explanation. Transformation of space
of differential forms induced by general linear coordinate transformation
is really valid representation of GL(4,$\R$), but that is relation
with usual spinor representation in case of restriction to
SO(3,1) $\subset$  GL(4,$\R$)? For example, exterior space in respect to
GL(4,$\R$) have five irreducible subspaces corresponding to spaces
$\Ext^k_4$, $k=0,\ldots,4$.

On the other hand, from a na\'{\i}ve point of view Dirac spinors in
such classification should correspond to some (fictitious) index like
$k=1/2$, because exterior form with index $k=1$ corresponds to (co)vector,
{\em i.e.} ``spin one.''

To justify the proposition, let us show first, that for Minkowski
metric and Lorentz group SO(3,1) $\subset$ GL(4,$\R$) suggested
transformation corresponds to
\begin{equation}
\nPsi \mapsto \LA \nPsi \LA^{-1}
\label{IntIs}
\end{equation}
used for description of transformation property \eq{LPsiL} of
Dirac equation in matrix representation \eq{nDir}.

Let us consider subspaces $\Cl_k(n) \subset \Cl(n)$, $k=0,\ldots,n$
produced by products of $k$ matrices $\gamma_\mu$. Only for element
$\LA$ from spin group internal isomorphism \eq{IntIs} of Clifford
algebra maps subspaces $\Cl_k(n)$ to itself. On the other hand, such
isomorphism may be expressed using substitution of generators like
\eq{Agam} with isometry $A$, {\em i.e.} Lorentz group for particular case
under consideration.

For Lorentz group transformation law for $\Cl_k(3,1)$ due to
\eq{Agam} is the same as for $\Ext_4^k$ and \eq{Adiff} respectively.
More formally, in such a case it is possible to express transformation
suggested in proposition as
\begin{equation}
 \bigl(\ClLGr(\LA_A)\circ\ClRGr(\LA_A^{-1})\bigr) (M),
 \quad M \in \Ext_4,~ \LA_A \in {\rm Spin}(3,1),~A \in {\rm SO}(3,1).
\end{equation}

\medskip

Where is also other way to represent constructions suggested above.
Let we have Dirac algebra of $4 \times 4$ complex matrices and some
fixed basis of Dirac matrices $\gamma_\mu$. Let us now together with
usual matrix multiplication introduce other associative and distributive
operation ``$\wedge$'' induced by structure of Grassmann algebra due to linear
map $\ExCl$ \eq{ExtToCl}. Formally, it would be necessary to write $16^2=256$
products for elements of basis, but really the operation ``$\wedge$'' is
unique defined by expression with two matrices
\begin{equation}
 \gamma_\mu \wedge \gamma_\mu \equiv 0,\quad
 \gamma_\mu \wedge \gamma_\nu \equiv -\gamma_\nu \wedge \gamma_\mu
 \equiv \gamma_\mu\gamma_\nu \quad (\mu < \nu),
\label{gamgrass}
\end{equation}
associativity and property $\gamma_{i_1} \wedge \cdots \wedge \gamma_{i_k}
 \equiv \gamma_{i_1} \cdots \gamma_{i_k}$ for $i_1< i_2 < \cdots < i_k$.

Now linear map \eq{Agam} with arbitrary  $A \in \rm GL(4,\R)$ may be extended
to representation of GL(4,$\R$) on full 16D space using ``$\wedge$'' products
of generators. It is also valid for $A \in \rm SO(3,1)$, but in such a case
difference between ``$\wedge$'' and usual matrix (Clifford) products does not
matter, because due to property of spin group usual product does not produce
``junk'' terms with $g_{\mu\nu}\Id$ (unit of algebra multiplied on some
coefficient of metric). The other property of $A \in \rm SO(3,1)$ is
possibility to express considered representation as internal isomorphism with
respect to usual product, $\nPsi \mapsto \LA_A \nPsi \LA_A^{-1}$, \eq{IntIs}.

It was already discussed earlier, how \eq{IntIs} corresponds to
usual spinor representation. Because of \eq{bipsi} matrix function $\nPsi$
may be associated with tensor product of usual spinor $\psi$ on some
``auxiliary spinor $\alpha$'' and then due to \eq{LMR}, it is possible
for Lorentz group to rewrite \eq{IntIs} as
\begin{equation}
 \psi\otimes\alpha \mapsto (\LA \psi) \otimes ({\LA^{-1}}^T \alpha),
\label{LocLor}
\end{equation}
but state of auxiliary system for such a product does not matter
due to possibility to apply arbitrary transformation
$R : \alpha \mapsto R^T \alpha$ (see Sec.~\ref{Sec:Cliff}).
So it is possible to take into account only transformation of
first term $\psi \mapsto \LA \psi$ in tensor product, {\em i.e.}
spinor representation of Lorentz group.

On the other hand, general GL(4,$\R$) transformation does not correspond to
map between ``product states'' like $\psi\otimes\alpha$. Using some
quantum mechanical jargon it could be possible to say, that general
GL(4,$\R$) transformation ``entangles'' state $\psi$ and auxiliary
state $\alpha$, and transformation $R$ on second state used above
may not improve situation (``disentangle'' states), because it
is ``local''. With same analogy, relation between $\nPsi$ and usual
Dirac spinor $\psi$ may be compared with conception
of {\em subsystem} in quantum mechanics.
But really such jargon should be considered only as
some hint, because more detailed consideration may use also real
representation of Clifford algebra and constructions
used above may correspond to real or even quaternionic matrices
and tensor products and, after all, transformations used here correspond
to some finite-dimensional unitary representations only for SO(3) subgroup of
GL(4,$\R$).

\end{document}